\documentclass[12pt,a4paper]{article}
\usepackage[utf8]{inputenc}
\usepackage{amsmath}
\usepackage{amsfonts}
\usepackage{amssymb}
\usepackage{mathtools}
\usepackage{graphicx}
\usepackage{cite}
\usepackage[capitalise]{cleveref}
\graphicspath{ {./figures/} }
\usepackage[left=2cm,right=2cm,top=2cm,bottom=2cm]{geometry}
\crefformat{section}{#2section~#1#3}

\author{Tetiana Obikhod and Ievgenii Petrenko}
\title{\bf{Searches for  $pp\rightarrow H^{\pm}h\rightarrow W^{\pm}hh$ signal in the MSSM model}}
\date{%
    {\it Institute for Nuclear Research NAS of Ukraine, Kyiv 03028, Ukraine}\\%
    \today
}
\begin{document}

\maketitle

\section{Abstract}

	We considered the new signature of a charged Higgs boson, $pp\rightarrow H^{\pm}h\rightarrow W^{\pm}hh$, which is almost background free as the promising to study of supersymmetry at the LHC with both $\sqrt{s} = 13$ TeV and $\sqrt{s}= 14$ TeV. In the framework of MSSM model we used the latest theoretical parameter space, which is consistent with the experimental constraints from Large Hadron Collider and direct or indirect searches. In the surviving regions with the help of Pythia 8.3 and Feynhiggs programs it was found that production cross sections, $\sigma(pp\rightarrow H^{\pm}h)$,  and Branching Ratios, $BR(H^{\pm}\rightarrow W^{\pm}h)$, are the largest at Benchmark Points 4 and 14 correspondingly.

\section{Introduction}

    Experimental searches for supersymmetry, associated with a number of key unresolved problems in the Standard Model (SM), are the most privileged channel for the search for new physics beyond the Standard Model (BSM). Searches for the extended Higgs boson sector as the lightest superparticles with masses up to 1 TeV are the most promising in this aspect. This is facilitated by both experimental data on searches for superparticles with masses of up to 100 GeV \cite{1., 2.} and theoretical constructions. Among supersymmetric models, the most popular is 2-Higgs Doublet Model (2HDM) with few Benchmark Points (BPs) which
pass all present constraints, both theoretical and experimental. There were performed Monte Carlo (MC) analysis of  different supersymmetry search channels, one of which, $W^{\pm} + 4\gamma$, is very promising \cite{3.}, as it can be connected with significant excesses at the LHC. 

    Since 2HDM is a tree-level Minimal supersymmetry standard model (MSSM) \cite{4.}, the purpose of the article is to carry out calculations for the production cross sections $\sigma(pp\rightarrow H^{\pm}h)$,  and Branching Ratios (BR) $BR(H^{\pm}\rightarrow W^{\pm}h)$ using the selected BP scenarios within the extended supersymmetry model, MSSM. It is important to select the most significant events based on the calculation results and find the corresponding masses and momentum distributions of the Higgs bosons. In addition, it is necessary to compare the obtained data with similar calculations within the framework of 2HDM in order to select the most suitable scenarios for the search for supersymmetry.

\section{Theory of 2HDM and MSSM model}
Searches for BSM physics, especially supersymmetry, is a privileged area of modern experimental high-energy physics. The most popular theory of supersymmetry MSSM.  At tree-level this model goes to another model, 2HDM as its low-energy limit with neutral CP-even scalars (h and H), CP-odd pseudoscalar (A) and two charged Higgs bosons (H$^{\pm}$). To show this, it must be noted that in the MSSM, the terms contributing to the scalar Higgs potential VH come from three different sources:\\
    i) the D terms
\[V_D=\frac{g^2_2}{8}\Biggl[4|H^+_1\cdot H_2|^2-2|H_1|^2|H_2|^2+(|H_1|^2)^2+(|H_2|^2)^2\Biggr]+\frac{g^2_1}{8}(|H_2|^2-|H_1|^2)^2\]
	ii) the F terms
\[V_F=\mu^2(|H_1|^2+|H_2|^2)\]
	iii) a piece originating from the soft SUSY–breaking scalar Higgs mass terms and the bilinear term
	\[V_{soft}=m^2_{H_1}H^+_1H_1+m^2_{H_2}H^+_2H_2+B\mu(H_2\cdot H_1+h.c.)\]
	The full scalar potential involving the Higgs fields is the sum of the three terms \cite{5.}
\[	V_H=(|\mu|^2+m^2_{H_1})|H_1|^2+(|\mu|^2+m^2_{H_2})|H_2|^2-\mu Be_{ij}(H^i_1H^j_2+h.c.)+\]
\begin{equation}
	\frac{g^2_2+g^2_1}{8}(|H_1|^2-|H_2|^2)^2+\frac{1}{2}g^2_2
	|H^+_1H_2|^2
\end{equation}
The most general supersymmetric potential for two Higgs doublets (H$_u$ and $H_d$) at tree-level reads \cite{6.}
\[V^{LO}_{MSSM}=(m^2_{H_d}+\mu^2)|H_d|^2+(m^2_{H_u}+\mu^2)|H_u|^2\]
\[-B\mu\varepsilon_{ij}(H^i_dH^j_u+h.c.)\]
\begin{equation}
+\frac{g^2+g^{'2}}{8}(|H_d|^2-|H_u|^2)^2+\frac{g^2}{2}|H^{*}_dH_u|^2
\end{equation}
where $m^2_{H_d},\ m^2_{H_u}, \  B\mu$  denote the soft-SUSY breaking mass parameters, $g$ and $g^{'}$ - the SU(2)$_L$ and U(1)$_Y$
gauge couplings. From comparison of formula (1) and (2) we see matching expressions, which confirms the coincidence of 2HDM with the MSSM model at the tree-level. The advantage of 2HDM model is the possibility to avoid large effects of Flavour Changing Neutral Currents. 
    
\section{Results of calculations} 
   Using article data  \cite{7.} we'll focus on the decay modes of the charged Higgs boson $H^{\pm}\rightarrow W^{\pm *} h$ in the 2HDM Type-I satisfying the condition $M_{H^{\pm}}< M_t + M_b$. It is assumed that the properties of heavy scalar H are consistent with the LHC measurements, and h is lighter than 125 GeV.  In this paper was emphasized that the production and decay process $pp\rightarrow H^{\pm}h\rightarrow W^{\pm *}hh\rightarrow l^{\pm}\nu +4\gamma$ is background free, so $W^{\pm} + 4\gamma$ signal would yield significant excesses at the LHC with an integrated luminosity of L = 300 fb$^{-1}$ at both $\sqrt{s}$ = 13 and 14 TeV.
 
     So, this search channel is of particular interest for exploiting a MC analysis at a detector level by including parton shower, hadronisation and heavy flavour decays, \cite{8.}, so that significances only depend upon the signal cross sections and the integrated luminosity. Moreover, the estimate of the signal significance over the Type-I parameter space, could be useful for LHC experimental groups. 
   
     To estimate the production cross section of considered signal $\sigma(pp\rightarrow H^{\pm}h)$ we choose the input parameters of more general MSSM model in the form of BP of  \cite{7.}, presented in Table 1   
\begin{center} 
{\it\normalsize Table 1. Input parameters of MSSM model}\\
\vspace*{3mm}
\begin{tabular}{|c|c|c|c|c|} \hline
BP&$M_h$&$M_A$&$M_{H^{\pm}}$&tan$\beta$ \\ \hline \hline
BP3&45.34&162.07&128.00&7.57 \\ \hline
BP4&53.59&126.09&91.49&8.00 \\ \hline
BP5&63.13&85.59&104.99&18.09 \\ \hline
BP6&65.43&111.43&142.15&11.52 \\ \hline
BP7&67.82&79.83&114.09&8.94 \\ \hline
BP9&73.18&108.69&97.34&8.06 \\ \hline
BP14&81.53&225.76&168.69&9.75 \\ \hline
\end{tabular}
\end{center}

Then with the help of Pythia 3 \cite{9.} we calculated production cross sections at 13 and 14 TeV presented in Table 2 and Table 3
\begin{center}
{\it\normalsize Table 2. Production cross sections, $\sigma(pp\rightarrow H^{\pm}h)$, at 13 TeV \\ for the corresponding BPs}\\
\vspace*{3mm}
\begin{tabular}{|c|c|c|} \hline 
BP& $\sigma$ (in pb) & statist. err.\\ \hline \hline
BP3&       5.235e-01   &    8.977e-03   \\ \hline
BP4 &      1.104e+00   &    1.877e-02   \\ \hline
BP5 &     6.715e-01   &    1.103e-02   \\ \hline
BP6 &    2.813e-01   &    4.740e-03   \\ \hline
BP7 &   6.287e-01   &    1.044e-02   \\ \hline
BP9 &   1.506e-01   &    2.536e-03   \\ \hline
BP14&   0.000e+00   &    0.000e+00   \\ \hline
\end{tabular}
\end{center}

\begin{center}
{\it\normalsize Table 3. Production cross sections, $\sigma(pp\rightarrow H^{\pm}h)$, at 14 TeV \\ for the corresponding BPs}\\
\vspace*{3mm}
\begin{tabular}{|c|c|c|} \hline 
BP& $\sigma$ (in pb) & statist. err.\\ \hline \hline
BP3  &     5.756e-01     &  9.561e-03   \\ \hline
BP4  &     1.232e+00     &  2.076e-02   \\ \hline
BP5  &     7.049e-01     &  1.161e-02   \\ \hline
BP6  &     3.278e-01     &  5.297e-03   \\ \hline
BP7  &     7.149e-01     &  1.190e-02   \\ \hline
BP9  &     1.668e-01     &  2.783e-03   \\ \hline
BP14 &     0.000e+00     &  0.000e+00   \\ \hline
\end{tabular}
\end{center}

Using FeynHiggs program \cite{10.} we calculated BR$(H^{\pm}\rightarrow W^{\pm}h)$, presented in Table 4. 
\begin{center}
{\it\normalsize Table 4. Branching Ratios, $BR(H^{\pm}\rightarrow W^{\pm}h)$, for the corresponding BPs}\\
\vspace*{3mm}
\begin{tabular}{|c|c|} \hline 
BP& BR\\ \hline \hline
BP3   & 0.002885294 \\ \hline 
BP4   & 0.001397527\\ \hline 
BP5   & 0.0005224388\\ \hline 
BP6   & 0.000561978\\ \hline 
BP7   & 0.00259184\\ \hline 
BP9   & 0.001528386\\ \hline 
BP14  & 0.02296814\\ \hline 
\end{tabular}
\end{center}

We received the mass and momentum distributions for the maximal value of production cross section of BP4 for h boson, Fig. 1

\begin{center}
\hspace*{-0.5cm} \includegraphics[width=0.5\textwidth]{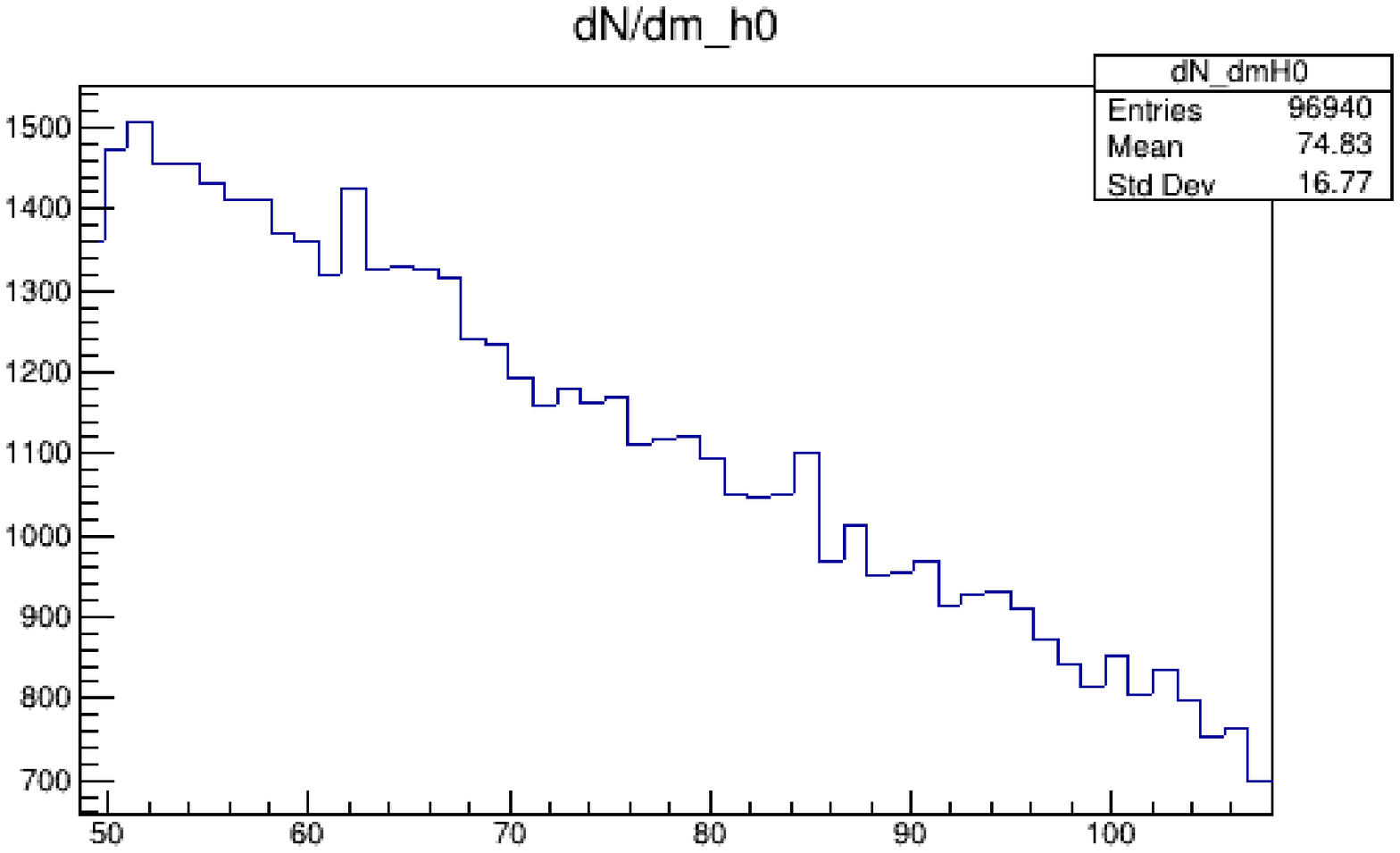}
\includegraphics[width=0.5\textwidth]{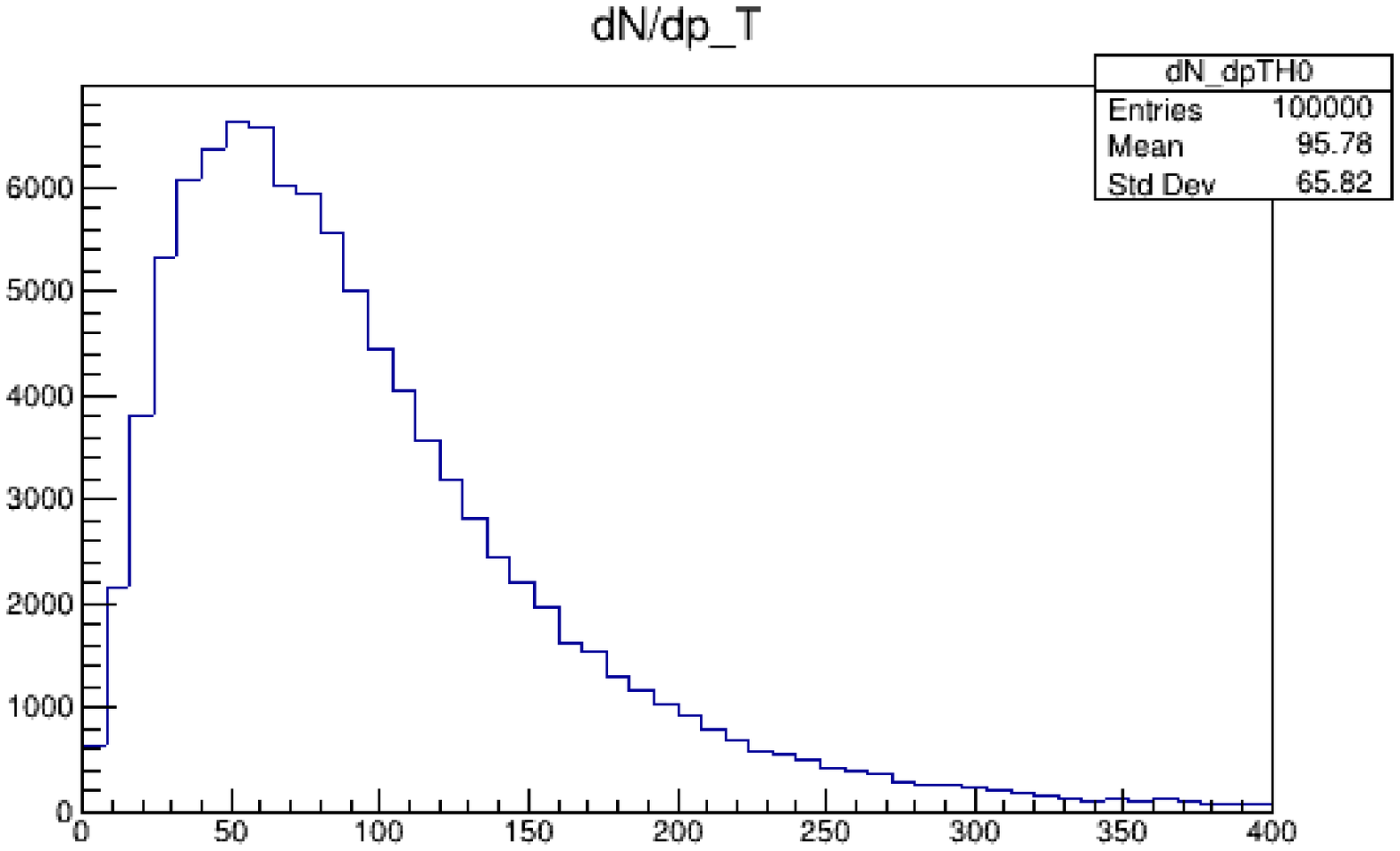}\\
\emph{\textbf{Fig.1}} {\emph{Mass (left) and momentum (right) distributions of h boson at BP4 scenario.  }}
\end{center}
As for H$^+$ boson we obtained the data in Fig. 2
\begin{center}
\hspace*{-0.5cm} \includegraphics[width=0.5\textwidth]{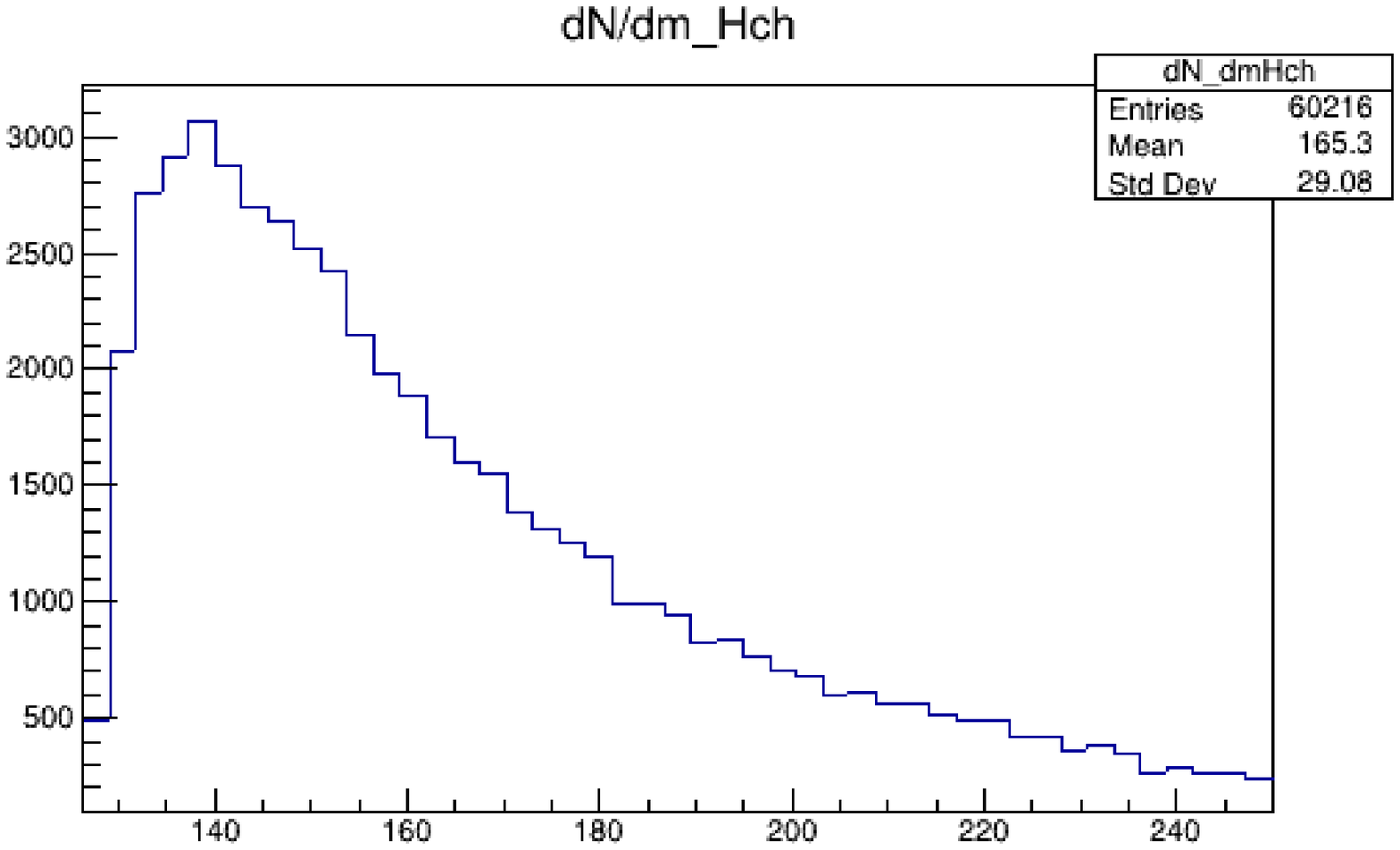}
\includegraphics[width=0.5\textwidth]{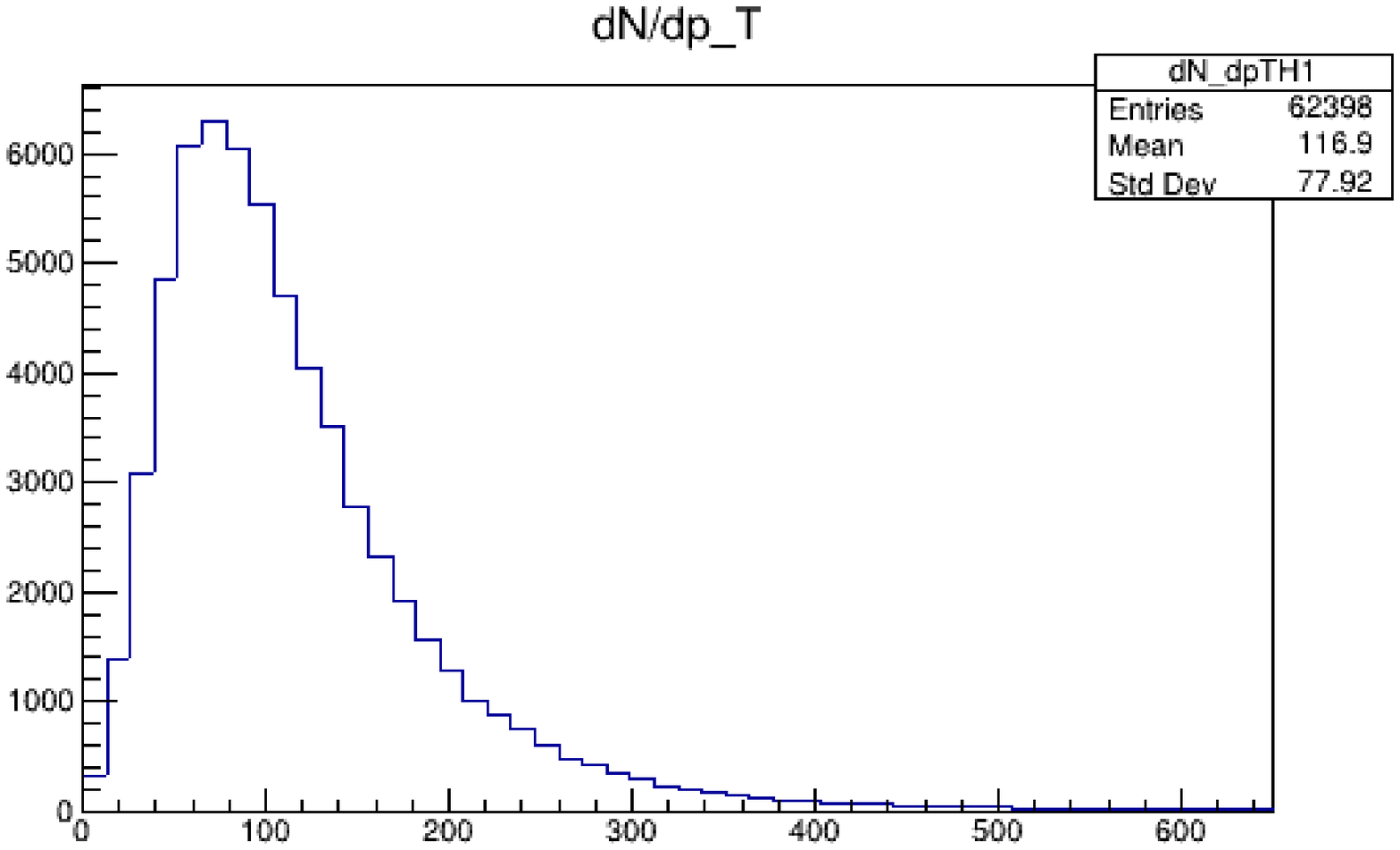}\\
\emph{\textbf{Fig.2}} {\emph{Mass (left) and momentum (right) distributions of H$^+$ boson at BP4 scenario.  }}
\end{center}

    From the data presented in Fig. 1 and Fig. 2, we can conclude that the mass of the lightest Higgs boson is 50 GeV at a maximum momentum of 50 GeV and the mass of a charged H$^+$ boson is 140 GeV at a maximum momentum of 100 GeV. As for BR, it is not changed with energy and has maximal value for BP14 parameter space without change of the M$_h$ and M$_{H^{\pm}}$ parameters and is equal to 0.02296814.

    It should be emphasized that the supersymmetric structure of the theory has imposed very strong constraints on the Higgs spectrum, so for the calculations we used six parameters which describe the MSSM Higgs sector, M$_h$, M$_H$, M$_A$, M$_{H^{\pm}}$, $\beta$, $\alpha$. Out of the six parameters which describe the MSSM Higgs sector, only two parameters, tan$\beta$ and M$_A$, are free parameters at the tree-level. In this case, calculations according to Pythia for two parameters, tan$\beta$ and M$_A$  give  almost identical results for any BP choice, which is presented in Table 5.
\begin{center}
{\it\normalsize Table 5. Production cross sections, $\sigma(pp\rightarrow H^{\pm}h)$, at 13 TeV \\ for the corresponding BPs at tree-level of MSSM model}\\
\vspace*{3mm}
\begin{tabular}{|c|c|c|} \hline 
BP& $\sigma$ (in pb) & statist. err.\\ \hline \hline
BP3  &     2.152e-02  &     3.688e-04 \\ \hline
BP4  &     2.090e-02  &     3.577e-04\\ \hline
BP5  &     2.065e-02  &     3.420e-04\\ \hline
BP6  &     2.128e-02  &     3.594e-04\\ \hline
BP7  &     2.106e-02  &     3.608e-04\\ \hline
BP9  &     2.124e-02  &     3.619e-04\\ \hline
BP14 &     0.000e+00  &     0.000e+00 \\ \hline
\end{tabular}
\end{center}

\section{Conclusions}

We compared MSSM and 2HDM models and demonstrated their coincidence at tree-level. We considered the restricted parameter space of more general MSSM model with six parameters, connected with the experimental constraints from Large Hadron Collider in the form of Benchmark Points. In the surviving regions with the help of Pythia 8.3 and Feynhiggs programs it was found that production cross sections, $\sigma(pp\rightarrow H^{\pm}h)$,  and $BR(H^{\pm}\rightarrow W^{\pm}h)$, are the largest at BP4 and BP14 correspondingly. As for the calculations within 2HDM model \cite{7.}, the largest values of production cross sections are observed for the BP8 scenario at both energies of 13 and 14 TeV. The corresponding mass distributions for the  lightest Higgs boson is 50 GeV at a maximum momentum of 50 GeV and the mass of a charged H$^+$  boson is 140 GeV at a maximum momentum of 100 GeV. As for BR, it has maximal value for BP14 parameter space without change of the M$_h$ and M$_{H^{\pm}}$ parameters and is equal to 0.02296814.

\end{document}